%% file: main.tex
\documentclass[11pt]{article}

\usepackage[a4paper,margin=1in]{geometry}
\usepackage[T1]{fontenc}
\usepackage{amsmath,amssymb}
\usepackage{booktabs}
\usepackage{graphicx}
\usepackage{microtype}
\usepackage[numbers,sort&compress]{natbib}
\usepackage{tabularx}
\usepackage{xcolor}
\usepackage{xspace}
\usepackage{enumitem}
\usepackage{caption}
\usepackage{tikz}
\usetikzlibrary{arrows.meta,positioning}
\usepackage[hidelinks]{hyperref}

\input{metadata.tex}

\newcommand{\system}{\textsc{FaultLens}\xspace}
\newcommand{\candidate}{\pi_c}

\newcommand{\nullpolicy}{\pi_0}

\setlist{nosep,leftmargin=1.5em}
\hypersetup{
  pdftitle={FaultLens: Learning Compact Behavioral Test Suites for Generated Operational Programs},
  pdfauthor={\PaperPDFAuthors}
}

\title{\system: Learning Compact Behavioral Test Suites\\
for Generated Operational Programs}
\author{\PaperAuthors\\[2pt]\small \PaperAffiliation}
\date{August 2026}

\begin{document}
\maketitle

\begin{abstract}
Generated operational programs are commonly validated with either a few
hand-written examples or an exhaustive regression suite. The former misses
sparse boundary and interaction faults; the latter can be unnecessarily
expensive when validation crosses time, observations, thresholds, and action
budgets. We introduce \system, a method for learning compact behavioral test
suites while preserving an auditable connection to executed evidence. The
method executes a rich probe domain once, stores the resulting fault--probe
kill relation as a sparse outcome cache, and learns probe orderings only from
earlier program generations. A fault-driven greedy component exploits known
kill structure. A mutation-independent diversity component covers probe
families, cases, templates, and temporal bins. Their alternating hybrid is
designed to remain useful when the next program contains a fault mechanism not
seen during ordering construction.

We study twenty generated operational policies in a resource-selection case
study with four environments, ten execution seeds, 1,200 measured run
summaries, 2,160 controlled program transformations, and 4,120,200 executed
program--probe pairs. Of 1,960 intended faulty transformations, 1,779 alter a
contract or output somewhere in the finite audit domain; 200 additional
controls preserve behavior. A 32-probe hybrid learned on program generations
1--3 covers 576/582 (99.0\%) dynamically killable faults in generations 4--5,
using 1.2--2.0\% of the exhaustive domain. When an entire fault family is
withheld from ordering construction, diversity raises scenario--family macro
coverage from 84.6\% to 94.9\%. As a downstream deployment study, a conservative
admission rule reduces severe tail regressions from 15/20 program--environment
groups to 0/20. The results show that compact test selection can generalize
across generated programs and partially across fault mechanisms, while also
making its misses explicit.
\end{abstract}

\section{Introduction}

Code-generating models increasingly produce small operational programs that
map telemetry and state to concrete actions. Examples include selecting a
cache policy, choosing a replica set, activating a feature configuration, or
allocating a limited set of database indexes. Such programs are attractive
because they can be inspected, versioned, tested, and rolled back. They are
also difficult to validate: a program can parse and return legal actions while
encoding a shifted phase boundary, an order dependency, stale state, a damaged
threshold, or an unsafe resource set.

Other generated artifacts expose similarly structured requirements for
physical consistency, temporal prompt transitions, and content--style
separation \cite{zheng2026vpt,tan2026swift,zhang2026sefs}.

Executing generated artifacts is now a standard part of agent evaluation
\cite{jimenez2024swebench}, but the validation budget is rarely treated as a
first-class object. A few examples provide weak behavioral evidence.
Exhaustive scenario-derived suites provide better evidence but may cross every
round, observation type, threshold neighborhood, multiplicity, and
permutation. That cost is especially visible when a probe invokes an external
planner or remote system rather than a microsecond-scale local function.

This paper asks a narrower question than program synthesis or policy learning:
\emph{given an interpretable exhaustive probe domain and historical generated
programs, can we learn a much smaller suite that retains fault sensitivity on
future programs?} The distinction matters. We do not generate the operational
program, optimize its task objective, or claim semantic correctness. We learn
which already-defined behavioral probes should run first under a declared
budget.

The main challenge is generalization. Mutation-guided prioritization can find
high-yield tests, but it can overfit to correlated mutants. A selector trained
on several threshold faults may repeatedly choose nearly identical threshold
probes. The next failure may instead be temporal or order-sensitive. \system
therefore combines two rankings:

\begin{enumerate}
  \item an \emph{active} ranking greedily covers training mutants using their
  executed kill relation; and
  \item a \emph{diversity} ranking ignores mutant outcomes and covers
  structural features of the probe domain.
\end{enumerate}

The hybrid alternates unique probes from both lists. This fixed allocation
spends half the nominal budget exploiting observed evidence and half exploring
scenario structure. We evaluate it with two separation protocols. The first
learns from program generations 1--3 and evaluates generations 4--5. The second
removes an entire fault family from training and evaluates only that family on
future programs.

Our contributions are:

\begin{itemize}
  \item a sparse counterfactual outcome cache that evaluates arbitrary probe
  subsets without synthesizing outcomes or rerunning programs;
  \item a hybrid active--diversity ordering for compact behavioral validation;
  \item cross-program and whole-fault-family protocols that expose selector
  overfitting; and
  \item an executed case study comprising twenty generated programs, 2,160
  transformations, 4.12 million program--probe executions, downstream
  operational measurements, and fully reproducible figures.
\end{itemize}

The strongest result is not exhaustive detection. The complete domain defines
which transformations are behaviorally effective. The substantive result is
that a ranking learned without the held-out programs retains 99.0\% coverage at
32 probes, and structural diversity materially improves generalization when
the held-out fault mechanism is unknown.

\section{Problem Setting}

\subsection{Generated Operational Policies}

At decision round $t$, an operational program receives observations $O_t$ and
returns an action set $A_t$ under resource budget $B$:
\begin{equation}
  \pi(t,O_t)\rightarrow A_t,
  \qquad A_t\subseteq\mathcal{A},\quad |A_t|\le B,
\end{equation}
where $\mathcal{A}$ is the declared action catalog. The program is expected to
be deterministic for identical inputs, invariant to duplicate records when
observations represent a set, and invariant to input permutation when order is
not part of the task semantics. These properties are domain contracts, not
universal properties of all software.

The validation input is a generated candidate $\candidate$ and a finite probe
domain $\mathcal{P}_s$ for environment $s$. Every probe records a round, an
observation multiset, a family, and a human-readable case. The output validator
normalizes action order and checks type, uniqueness, budget, and catalog
membership.

\subsection{Fault--Probe Relation}

Let $\mathcal{M}_s$ be controlled transformations of historical programs. A
probe $p$ kills mutant $m$ when the normalized output differs from the
unmodified program or becomes invalid:
\begin{equation}
  K(m,p)=\mathbf{1}\!\left[
  \operatorname{out}(m,p)\ne\operatorname{out}(\pi,p)
  \right].
\end{equation}
Static contract changes are tracked separately. A transformed program is
\emph{domain-effective} if it changes a static contract or an output somewhere
in the complete finite domain. Domain-equivalent mutants are reported and
excluded from dynamic-coverage denominators, following standard mutation-study
practice \cite{papadakis2019mutation}.

For a selected probe set $Q\subseteq\mathcal{P}_s$, dynamic coverage is
\begin{equation}
\label{eq:coverage}
 C(Q)=\frac{1}{|\mathcal{M}'_s|}\sum_{m\in\mathcal{M}'_s}
 \mathbf{1}\!\left[Q\cap\{p:K(m,p)=1\}\ne\emptyset\right],
\end{equation}
where $\mathcal{M}'_s$ is the evaluation universe of dynamically killable
mutants. The selector must choose an ordered prefix under budget $b$ using only
training programs and allowed training fault families.

\subsection{Evaluation Questions}

We ask four questions:

\begin{description}
  \item[RQ1.] Which probe families expose each controlled fault family?
  \item[RQ2.] How much cross-program fault coverage remains as the probe budget
  shrinks?
  \item[RQ3.] Does structural diversity help when an entire fault family is
  absent from training?
  \item[RQ4.] Can compact behavioral validation support a safer downstream
  deployment decision?
\end{description}

\section{Method}

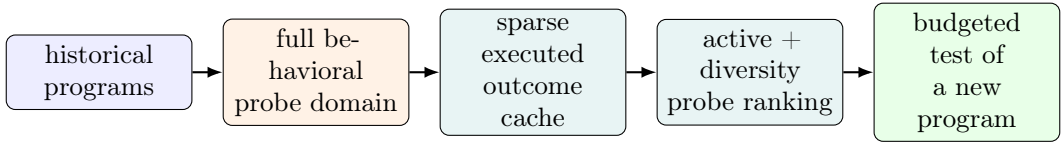
\begin{figure}[t]
\centering
\begin{tikzpicture}[
  node distance=4mm and 4mm,
  box/.style={draw,rounded corners=1.2mm,align=center,minimum height=8mm,text width=22mm,font=\small},
  arr/.style={-{Latex[length=2mm]},thick}
]
\node[box,fill=blue!7] (programs) {historical\\programs};
\node[box,fill=orange!10,right=of programs] (domain) {full behavioral\\probe domain};
\node[box,fill=teal!9,right=of domain] (cache) {sparse executed\\outcome cache};
\node[box,fill=teal!9,right=of cache] (rank) {active + diversity\\probe ranking};
\node[box,fill=green!9,right=of rank] (new) {budgeted test of\\a new program};
\draw[arr] (programs) -- (domain);
\draw[arr] (domain) -- (cache);
\draw[arr] (cache) -- (rank);
\draw[arr] (rank) -- (new);
\end{tikzpicture}
\caption{\system learns a compact ordering from executed historical evidence.
The new program executes the selected prefix rather than replaying cached
outputs.}
\label{fig:pipeline}
\end{figure}

Figure~\ref{fig:pipeline} summarizes the workflow. Historical programs are used
offline to construct executed outcome evidence and a probe ranking. A new
program still executes the selected probes; the cache does not predict its
outputs. Task measurements, when available, are a separate downstream
admission layer.

\subsection{Contracts Before Prioritization}

Test prioritization is useful only after inexpensive fail-closed checks. Our
case-study implementation parses Python into an AST, verifies one declared
entry point, rejects imports and dynamic call targets, and disallows global or
nonlocal mutation. Accepted code executes in a fresh process under CPU,
address-space, output-size, and wall-time limits. Return values are checked
again for type, duplicates, resource budget, and catalog membership.

These checks reduce accidental failures; they are not an adversarial sandbox.
The prioritizer operates only on baseline programs that pass the static and
metamorphic contracts. Static-only faults do not consume a dynamic probe
budget.

\subsection{Scenario-Derived Probe Domain}

The exhaustive domain crosses time with observation structure. At every round
it includes:

\begin{itemize}
  \item two identical empty inputs for repeatability;
  \item eight values around each declared threshold: $0$, $0.5$, $0.95$,
  $0.999$, $1.0$, $1.001$, $1.05$, and $2.0$ times the threshold;
  \item duplicate and triplicate observations for every template;
  \item both orders of every distinct template pair;
  \item mixed high--low template pairs; and
  \item an unknown observation template.
\end{itemize}

The resulting domains contain 1,620 probes for two-template environments and
2,700 probes for the three-template environment. Repeat, duplicate, and
permutation pairs also define metamorphic relations \cite{chen2018metamorphic}.
The grammar is intentionally interpretable: a selected probe maps back to a
round, threshold case, template combination, and declared relation.

\subsection{Sparse Executed Outcome Cache}

The full domain is executed once for every baseline and transformed program.
Instead of storing a dense $|\mathcal{M}|\times|\mathcal{P}|$ matrix, the cache
stores one record per mutant with static outcomes and only the IDs of killing
probes. Equation~\ref{eq:coverage} can then be evaluated by set intersection.

This representation enables thousands of counterfactual subset evaluations
without further program or system execution. It is exact relative to the
frozen full-suite trace; it does not establish equivalence under arbitrary
probe reorderings for a hidden-state program. Static state rejection and
repeatability relations reduce this concern, while high-assurance deployments
can execute each selected probe in a fresh process.

\subsection{Active Greedy Ranking}

For training mutant set $U$, the active selector repeatedly chooses
\begin{equation}
 p^*=\arg\max_{p\in\mathcal{P}\setminus Q}
 \left|\{m\in U:K(m,p)=1\}\right|,
\end{equation}
adds $p^*$ to $Q$, and removes its newly killed mutants from $U$. Ties are
broken by stable probe ID. This is a set-cover heuristic
\cite{chvatal1979setcover} and accounts for overlap ignored by sorting probes
on individual kill frequency.

Pure greedy selection can nevertheless over-specialize. Once it finds a
high-yield region, its early ranks can be dominated by structurally similar
probes. This is desirable exploitation when future faults match training, but
fragile when the fault mechanism shifts.

\subsection{Mutation-Independent Diversity Ranking}

The diversity ranking never reads $K$. It maps each probe to categorical
features:

\begin{itemize}
  \item probe family and case;
  \item five-round temporal bin;
  \item family--time and case--time interactions; and
  \item nonnumeric template tokens encoded by the stable probe ID.
\end{itemize}

A second greedy pass covers the largest number of unseen structural features.
The hybrid alternates unique probes from the active and diversity rankings.
The 1:1 ratio is fixed before evaluation. At very small budgets, the hybrid is
expected to trail active selection because some choices explore rather than
kill known mutants. At larger budgets, that exploration may improve transfer
to unseen programs and unseen fault families.

\section{Study Design}

\subsection{Operational Resource-Selection Case Study}

We evaluate generated functions that select a limited action set over 60
rounds. The concrete action is activating database indexes, but the validation
method sees only rounds, observation records, action catalogs, and budgets.
The four environments represent incident bursts, a three-phase composite
workload, periodic switching, and rare high-cost events. Each environment uses
a 120k-row SQLite instance.

Five independent generation runs produce one Python policy per environment,
yielding twenty program--environment groups. Ten execution seeds and six
operational baselines yield 1,200 frozen measurement rows. Program generation
is not evaluated in this paper; the generated artifacts are treated as fixed
subjects identified by hashes.

\begin{table}[t]
\centering
\caption{Case-study environments and full behavioral domains.}
\label{tab:domains}
\begin{tabular}{lrrrrr}
\toprule
Environment & Templates & Actions & Budget & Rounds & Probes \\
\midrule
Burst     & 2 & 5 & 2 & 60 & 1,620 \\
Composite & 3 & 7 & 1 & 60 & 2,700 \\
Cycle     & 2 & 4 & 1 & 60 & 1,620 \\
Rare      & 2 & 5 & 1 & 60 & 1,620 \\
\bottomrule
\end{tabular}
\end{table}

\subsection{Transformation Corpus}

Each program receives 98 intended faulty transformations across thirteen
families and ten behavior-preserving controls across five families. Fault
parameters vary temporal modulus and offset, active windows, round shifts,
templates, thresholds, and replacement actions.

\begin{table}[t]
\centering
\caption{Controlled transformations. ``Effective'' means a static contract or
output changes somewhere in the complete finite domain.}
\label{tab:mutations}
\begin{tabularx}{\linewidth}{@{}lrrX@{}}
\toprule
Category & Generated & Effective & Families \\
\midrule
Contract integrity & 440 & 440 & import, state, catalog, budget \\
Temporal behavior & 560 & 488 & dropout, shift, partial schedule \\
Observations & 720 & 621 & drop, duplicate, order, threshold, template \\
Action choice & 240 & 230 & action substitution \\
\midrule
Fault candidates & 1,960 & 1,779 & 13 families \\
Equivalent controls & 200 & 200 & 5 control families \\
\bottomrule
\end{tabularx}
\end{table}

The complete execution contains 4,120,200 program--probe pairs. Of 1,960 fault
candidates, 181 are equivalent on the finite domain, leaving 1,779 effective
instances. Sixty are detected only statically, 57 through both channels, and
1,662 only dynamically. None of the 200 controls changes a static contract or
output.

The complete-domain numbers establish the replay universe. They should not be
read as an independent 100\% detection claim: a dynamically effective mutant
is defined by changing at least one full-domain output.

\subsection{Generalization Protocols}

For \emph{cross-program} evaluation, generations 1--3 construct a separate
ranking for each environment and generations 4--5 are used only for evaluation.
Only contract-passing baseline programs enter the dynamic experiment. The
held-out universe contains 582 dynamically killable instances.

For \emph{whole-family holdout}, one fault family is removed from ordering
construction. The ranking is evaluated on that family in future programs. We
report macro coverage over 42 nonempty environment--fault-family cells, so
large mutation families do not dominate.

Baselines are active greedy, the hybrid, individual kill frequency, and 200
deterministic random rankings per environment. Random curves report means and
5th--95th percentile envelopes.

\subsection{Downstream Operational Admission}

Behavioral validity does not imply operational utility. To study downstream
use, each program is paired with a validated restricted reference policy.
Seeds 1--5 are used for admission and seeds 6--10 are held out. For execution
seed $i$, normalized loss is
\begin{equation}
 L_i(\pi)=\frac{1}{2}\frac{P_i(\pi)}{P_i(\nullpolicy)}+
 \frac{1}{2}\frac{V_i(\pi)}{\max(1,V_i(\nullpolicy))},
\end{equation}
where $P$ is p95 latency and $V$ is the number of operational threshold
violations. A free program is admissible only when it passes behavioral
contracts, improves calibration loss, and one-sided 95\% bootstrap upper bounds
on latency and violation harm satisfy a declared operating profile. The Safe
profile permits at most 5\% p95 harm; Balanced permits 30\%. Both permit 0.02
normalized violation harm. These profiles are fixed before held-out evaluation.

\section{Results}

\subsection{RQ1: Evidence Is Distributed Across Probe Families}

\begin{figure}[t]
\centering
\includegraphics[width=\linewidth]{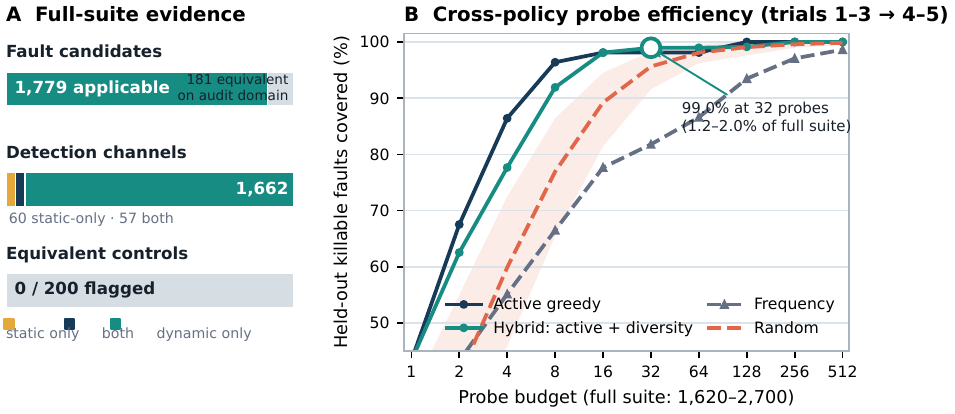}
\caption{Complete executed evidence and cross-program probe efficiency. The
ranking is learned on generations 1--3 and evaluated on generations 4--5.
Random shading is the weighted 5th--95th percentile range over 200 rankings.}
\label{fig:evidence}
\end{figure}

\begin{figure}[t]
\centering
\includegraphics[width=0.96\linewidth]{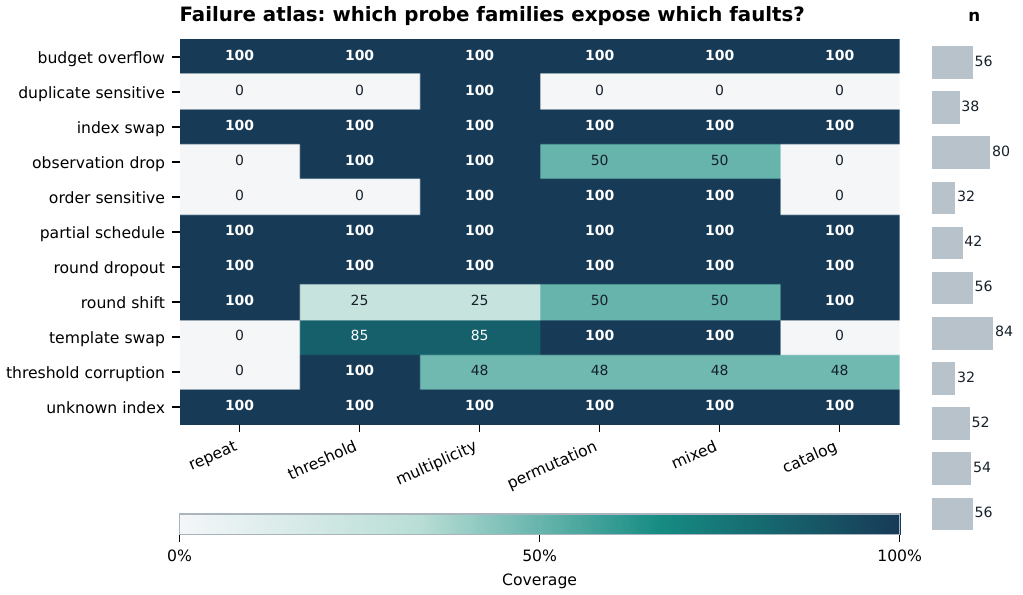}
\caption{Held-out failure atlas. Each cell gives the fraction of dynamically
killable future-program mutants in a fault row exposed by all probes in the
column family.}
\label{fig:atlas}
\end{figure}

Figure~\ref{fig:evidence}A summarizes the complete executed evidence. The
dynamic channel is essential: 1,662 effective faults do not change a static
contract. Figure~\ref{fig:atlas} provides the more informative view. Some
faults, including budget overflow and complete round dropout, are exposed by
nearly every probe family. Other rows are selective. Duplicate-sensitive
programs require multiplicity probes. Threshold corruption requires threshold
coverage. Order-sensitive programs require multi-observation inputs. Round
shifts combine temporal and catalog evidence.

No individual probe family dominates all fault rows. This motivates both a
rich complete domain and explicit structural exploration in the compact
ranking.

\subsection{RQ2: Compact Suites Transfer Across Programs}

Table~\ref{tab:budgets} reports scenario-weighted coverage. Active greedy is
strongest at budgets 4 and 8 because every early selection exploits observed
kills. At 16 probes, active and hybrid both cover 571/582 (98.1\%) held-out
faults, versus 89.2\% for random. At 32 probes, hybrid reaches 576/582 (99.0\%)
while using only 1.2\% of the 2,700-probe domain or 2.0\% of a 1,620-probe
domain. Composite, Cycle, and Rare reach 100\%; the six remaining misses occur
in Burst.

\begin{table}[t]
\centering
\caption{Cross-program held-out dynamic coverage (\%).}
\label{tab:budgets}
\begin{tabular}{rrrrr}
\toprule
Budget & Active & Hybrid & Frequency & Random \\
\midrule
4   & 86.4 & 77.7 & 55.2 & 59.9 \\
8   & 96.4 & 91.9 & 66.5 & 76.9 \\
16  & 98.1 & 98.1 & 77.7 & 89.2 \\
32  & 98.1 & 99.0 & 81.8 & 95.6 \\
64  & 98.1 & 99.0 & 86.6 & 98.1 \\
128 & 100.0 & 99.1 & 93.5 & 99.1 \\
\bottomrule
\end{tabular}
\end{table}

The result also shows that hybrid is not uniformly superior. Exploration costs
coverage below 16 probes. At 128 probes, active reaches 100\% while the
alternating hybrid remains at 99.1\%. A deployment with an extremely small
budget should prefer exploitation when the historical fault model is trusted;
a higher-assurance deployment can run a longer active prefix or the full
domain.

The full batch audit has median wall time 59.1 ms per baseline program, while
measured policy calls account for roughly 1.5 ms; process startup dominates in
this micro-policy case study. We therefore claim a reduction in executed probe
count and validation payload, not a measured 50$\times$ wall-time speedup.
Savings should be larger when probes invoke planners, containers, or remote
systems.

\subsection{RQ3: Diversity Helps on Unseen Fault Families}

\begin{figure}[t]
\centering
\includegraphics[width=0.92\linewidth]{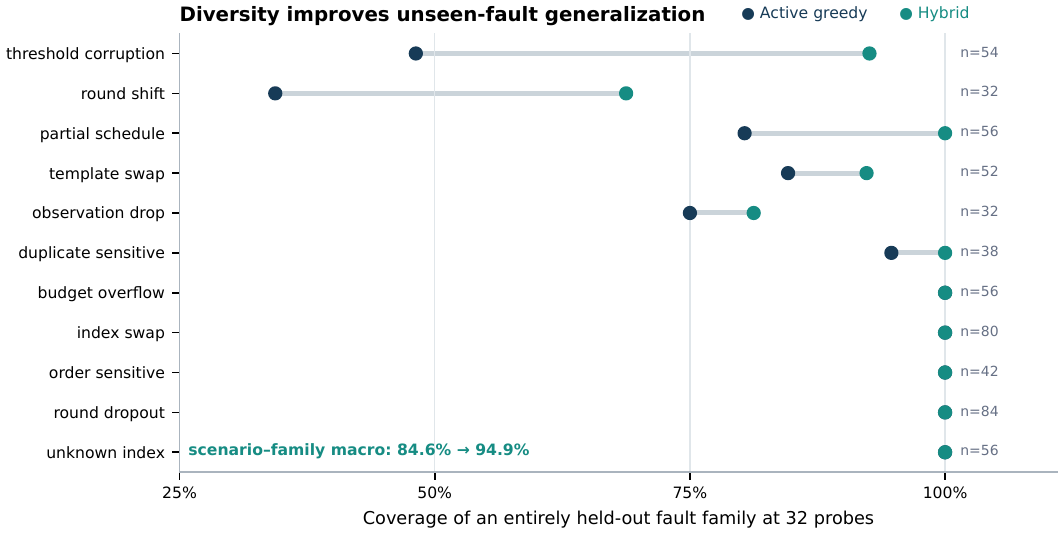}
\caption{Whole-fault-family holdout at budget 32. Ordering construction excludes
the row's fault family on generations 1--3; evaluation uses only that family in
generations 4--5.}
\label{fig:operator}
\end{figure}

Whole-family holdout creates a stronger distribution shift. At budget 32, pure
active greedy achieves 84.6\% macro coverage over the 42 nonempty
environment--family cells. Hybrid reaches 94.9\%. Figure~\ref{fig:operator}
aggregates by fault family. Diversity is equal or better on all eleven dynamic
families. The largest gains occur for threshold corruption, round shifts, and
partial schedules---precisely the faults that require coverage across
structural cases or temporal bins rather than another high-frequency training
probe.

The hybrid is not complete under this shift. Coverage is 68.8\% for round
shifts and 81.2\% for observation drop. Diversity is therefore a robustness
mechanism, not a substitute for adding incident-derived faults and probes to
the domain.

\subsection{RQ4: Downstream Admission Reduces Tail Regressions}

Table~\ref{tab:deployment} evaluates held-out seeds. Deploying every free
program yields normalized loss 0.0838 and severe p95 regressions in 15 of 20
program--environment groups. Always using the restricted reference yields
0.0269 and zero regressions. Selecting by calibration mean lowers loss to
0.0247 but accepts five regression groups.

The Safe profile selects two free programs, achieves loss 0.0269, and has zero
observed severe regressions. Balanced selects six, reduces mean violations from
10.33 to 6.19, and lowers loss to 0.0251 while accepting four regression groups.
The evaluation-only oracle also selects five regressing groups because scalar
loss can exchange large tail harm for violation reduction.

\begin{table}[t]
\centering
\caption{Held-out operational results. A severe regression is a
program--environment group with mean p95 more than 5\% above the reference.}
\label{tab:deployment}
\small
\begin{tabular}{lrrrr}
\toprule
Method & Loss & p95 ms & Violations & Regressions \\
\midrule
Free programs & 0.0838 & 0.852 & 29.15 & 15/20 \\
Restricted reference & 0.0269 & 0.303 & 10.33 & 0/20 \\
Calibration mean & 0.0247 & 0.357 & 5.16 & 5/20 \\
FaultLens-Safe & 0.0269 & 0.303 & 10.33 & 0/20 \\
FaultLens-Balanced & 0.0251 & 0.345 & 6.19 & 4/20 \\
Evaluation oracle & 0.0246 & 0.356 & 5.15 & 5/20 \\
\bottomrule
\end{tabular}
\end{table}

\begin{figure}[t]
\centering
\includegraphics[width=\linewidth]{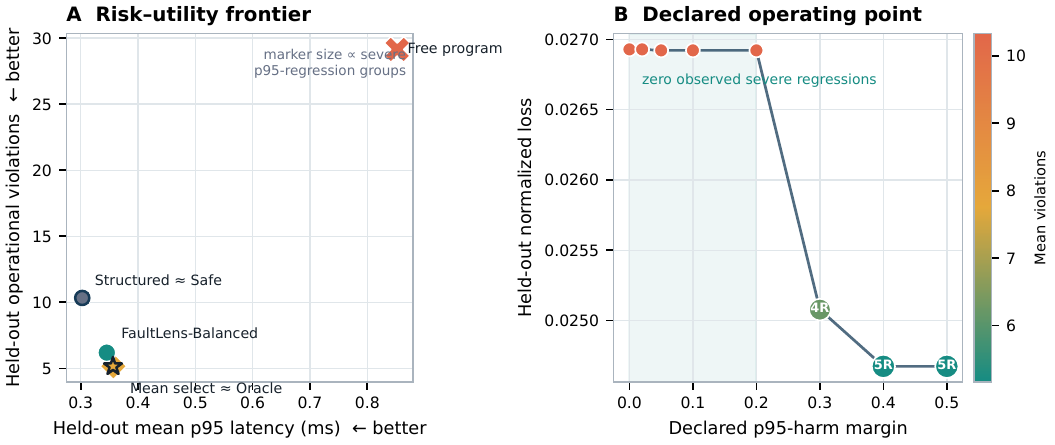}
\caption{Downstream risk and utility. Left: deployment strategies in the
latency--violation plane. Right: the declared p95-harm margin sweep; marker
labels give severe regression counts.}
\label{fig:risk}
\end{figure}

Figure~\ref{fig:risk} separates latency and violations. The declared p95-harm
margin creates a visible transition: margins through 0.20 have zero observed
severe regressions; at 0.30, four appear. A single scalar score would hide this
operating choice.

\section{Analysis and Ablations}

\subsection{Why Frequency Is Not Enough}

Frequency prioritization chooses probes with many individual training kills
but does not account for redundant coverage. At budget 32 it reaches 81.8\%,
compared with 98.1\% for active set cover. The 16.3-point difference measures
the value of updating the uncovered set after every selection. Test
prioritization work has long emphasized early fault revelation
\cite{rothermel1999prioritization,yoo2012regression}; generated programs add the
need to define a scenario grammar and an explicit program-generation split.

\subsection{Why Diversity Helps}

On the ordinary cross-program split, active and hybrid differ by less than one
point at budget 32. The whole-family holdout reveals a 10.3-point macro gap.
The diversity half of the ranking is therefore not primarily protecting
against new source syntax; it protects against a change in the relationship
between fault mechanism and informative probe structure.

This conclusion is limited to the chosen structural features. Five-round bins,
probe cases, and template tokens are natural for the case study but not
universal. Other operational programs may require topology, permission, API,
or concurrency features.

\subsection{Layer Contributions}

Static contracts alone detect 117 effective transformations but miss 1,662
output-only faults. Behavioral validation alone admits seventeen of twenty
baseline programs, yet the conservative downstream rule admits only two.
Conversely, statistical task measurements cannot reject imports, unknown
actions, hidden state, or order dependence. The layers answer different
questions:

\begin{itemize}
  \item contracts establish admissible code and output structure;
  \item compact probes establish behavioral evidence under a cost budget; and
  \item task measurements establish whether a valid program should be deployed
  at a declared risk operating point.
\end{itemize}

\subsection{Outcome Cache}

The sparse cache is 27 MiB for 2,160 transformations and 4,120,200 executed
program--probe pairs. Recomputing every budget curve, 200 random rankings per
environment, whole-family holdouts, and the failure atlas requires no program
or database execution. Each aggregate point still maps to concrete killing
probe IDs, avoiding a learned surrogate whose predictions would themselves
require validation.

\section{Limitations}

\textbf{Program domain.} The study covers twenty short deterministic Python
policies and four environments. Programs that call tools, manipulate schemas,
coordinate distributed components, or use concurrency need stronger isolation
and different structural features.

\textbf{Mutation representativeness.} Controlled transformations measure
sensitivity, not real-world fault prevalence. The thirteen families may favor
the grammar that motivated them. Cross-program splitting prevents direct
source memorization, and whole-family holdout reduces operator leakage, but
neither replaces a corpus of real incidents.

\textbf{Finite-domain equivalence.} The 181 excluded fault candidates are
equivalent only on the complete audit domain, not necessarily for every
possible input. Likewise, the 200 behavior-preserving controls cover comments,
whitespace, dead branches, identity transforms, and input copies rather than
all semantics-preserving transformations.

\textbf{State and ordering.} Cached subset coverage is exact for frozen
full-domain executions. It may not match a different execution order for a
hidden-state program. The method rejects explicit state and tests repeatability;
this is not a formal purity proof.

\textbf{Operational statistics.} The downstream study uses five calibration
and five held-out seeds per group. Bootstrap bounds cannot correct deployment
distribution shift, and the 5\% severe-regression threshold is an operational
definition rather than a theorem. Stronger distributional guarantees require
additional assumptions, as in conformal risk control
\cite{angelopoulos2024conformalrisk}.

\textbf{Security.} AST restrictions and resource limits reduce accidental
damage but do not safely execute adversarial code. Untrusted deployment requires
OS isolation, credential separation, syscall controls, and security review.

\section{Related Work}

\textbf{Property and behavioral testing.} QuickCheck popularized generated
property tests \cite{claessen2000quickcheck}; metamorphic testing derives
relations when individual outputs lack complete oracles
\cite{chen2018metamorphic}; CheckList organizes behavioral capabilities for
model evaluation \cite{ribeiro2020checklist}. \system uses a finite,
scenario-derived grammar and learns an ordering over its probes.

\textbf{Mutation testing.} Mutation analysis evaluates whether a suite exposes
controlled faults \cite{papadakis2019mutation}. Mutation-driven generation can
also construct tests and oracles \cite{fraser2011mutation}. Our full-domain
executions establish a sparse kill relation; program and operator holdouts
evaluate whether a compact subset transfers beyond the transformations used to
prioritize it.

\textbf{Regression-test prioritization.} Prior work orders regression tests to
reveal faults earlier \cite{rothermel1999prioritization,yoo2012regression}. The
active component is a deterministic set-cover heuristic. The diversity
component addresses correlated mutation evidence by reserving budget for
structural exploration.

\textbf{Operational learning and risk.} Learned database components often use
fallback mechanisms because single harmful decisions can dominate average
benefit \cite{marcus2021bao,vanaken2017ottertune}. Tail latency is especially
important in operational systems \cite{dean2013tail}. Bootstrap intervals
quantify sampling uncertainty \cite{efron1994bootstrap}; they do not provide
distribution-free deployment guarantees. Recent work on agent-trajectory
monitoring likewise uses an inexpensive calibrated triage layer to escalate
ambiguous or adversarial cases to stronger judges \cite{liu2026benchguard}.
\system keeps compact behavioral evidence separate from the downstream
performance rule.

\section{Reproducibility}

The evaluation artifact records stable probe IDs, transformation parameters,
static outcomes, sparse killing-probe lists, program hashes, deterministic
random seeds, bootstrap labels, and every plotted CSV row. The complete source
run executes 4,120,200 program--probe pairs. Cached replay reconstructs the
probe-budget curves and whole-family holdouts without executing programs or the
case-study system. The manuscript figures are vector PDFs and the present
arXiv source has no paths outside its upload directory.

Reproduction should distinguish two targets. A full run reconstructs executed
outcomes from frozen program sources. A replay run treats those outcomes as
immutable evidence and reconstructs selection experiments. This separation
makes it possible to audit prioritization logic without trusting a synthetic
performance model.

\section{Conclusion}

Generated operational programs need more evidence than a parser and a few
examples, but exhaustive validation is not always the right default. \system
learns compact behavioral test suites from executed historical evidence while
reserving part of the budget for mutation-independent structural diversity. In
the case study, 32 probes retain 99.0\% fault coverage across future programs,
and diversity improves whole-fault-family macro coverage from 84.6\% to 94.9\%.
The remaining misses are as important as the gains: compact validation is a
prioritized evidence mechanism, not a proof of correctness. Its value is that
the budget, evidence source, generalization split, and deployment operating
point are all explicit and auditable.

\appendix

\section{Per-Environment Budget Curves}

\begin{figure}[ht]
\centering
\includegraphics[width=\linewidth]{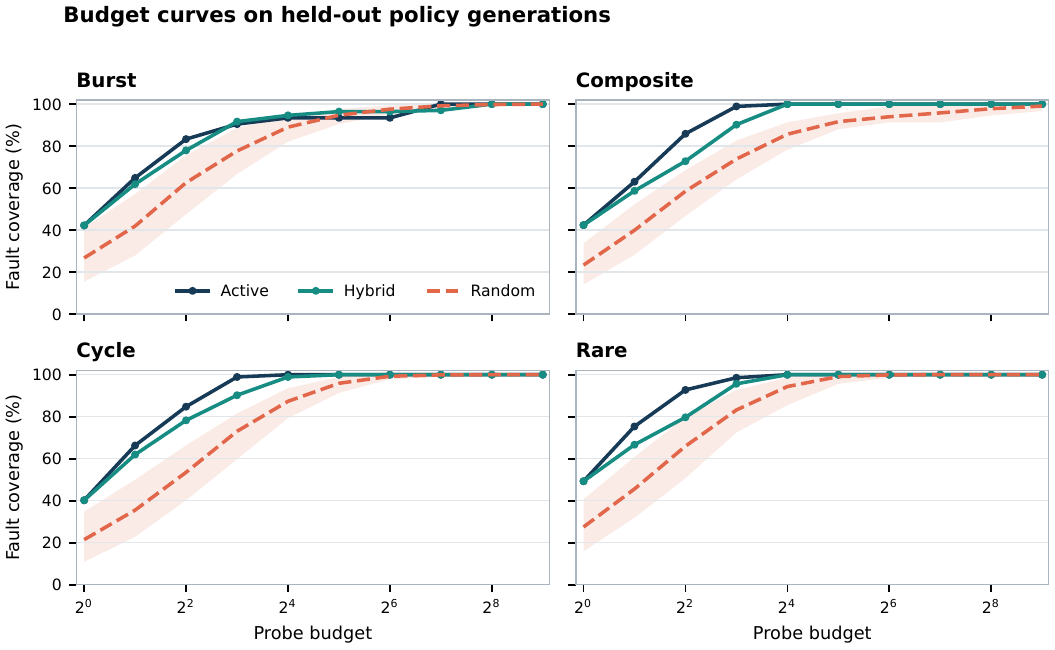}
\caption{Cross-program probe-budget curves for each evaluation environment.}
\label{fig:scenarios}
\end{figure}

Figure~\ref{fig:scenarios} expands the weighted curve into four environments.
The hybrid reaches 100\% by 16 probes in Composite, Cycle, and Rare. Burst
requires a longer prefix and accounts for all six cross-program misses at
budget 32. Reporting the weighted aggregate alone would conceal this
heterogeneity.

\section{Policy-Level Admission Landscape}

\begin{figure}[ht]
\centering
\includegraphics[width=0.82\linewidth]{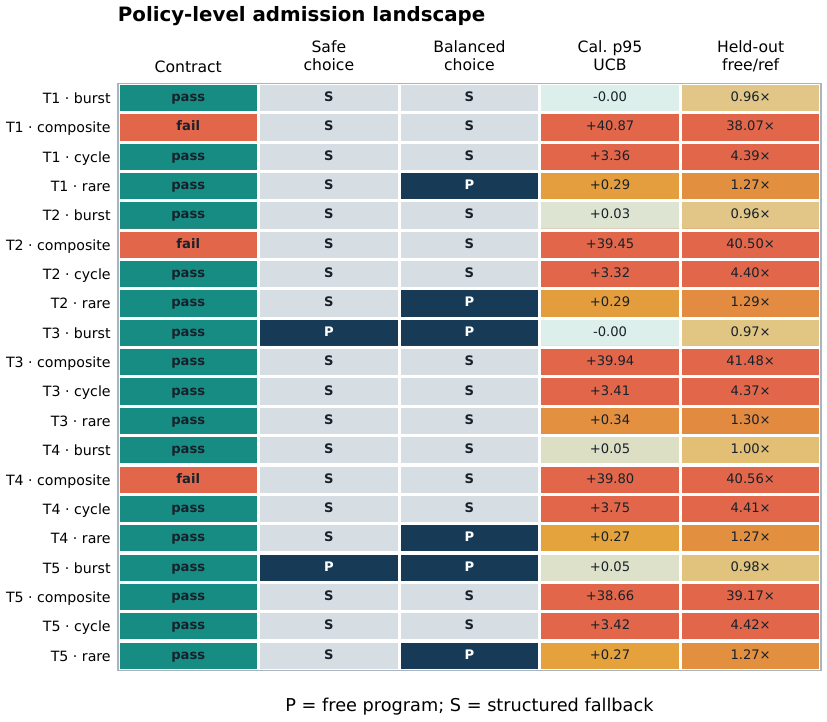}
\caption{Program-level downstream admission evidence.}
\label{fig:landscape}
\end{figure}

Figure~\ref{fig:landscape} shows all twenty downstream decisions. P denotes the
free generated program and S the restricted reference. The free programs in
Composite and Cycle often exhibit extremely large calibration upper bounds and
held-out tail ratios, whereas several Burst programs improve tail latency. The
aggregate result therefore does not arise from a uniform reject-all policy.

\begingroup
\small
\bibliographystyle{unsrtnat}
\bibliography{references}
\endgroup

\end{document}

%% file: metadata.tex
\newcommand{\PaperAuthors}{%
  Zeming Liu \qquad Hang Lyu \qquad Jingtao Zhang}
\newcommand{\PaperPDFAuthors}{Zeming Liu; Hang Lyu; Jingtao Zhang}
\newcommand{\PaperAffiliation}{Independent Researchers}